\documentclass[useAMS,usenatbib]{mn2e}
\usepackage{natbibmnfix}
\usepackage{graphicx}
\usepackage{grffile}
\usepackage{subfigure}
\usepackage{mathrsfs,amssymb}
\usepackage{amsmath}
\usepackage{natbib}
\usepackage{color}
\usepackage{ulem}
\usepackage{url}
\usepackage{bm}
\usepackage{multirow}
\usepackage[flushleft]{threeparttable}

\newcommand{\apj}{\rm {ApJ}}                 
\newcommand{\apjl}{\rm {ApJ}}                
\newcommand{\apjs}{\rm {ApJS}}               
\newcommand{\aap}{\rm {A\&A}}                

\newcommand{\mnras}{\rm {MNRAS}}             
\newcommand{\nat}{\rm {Nature}}              

\def\gtsima{$\; \buildrel > \over \sim \;$}
\def\ltsima{$\; \buildrel < \over \sim \;$}
\def\gsim{\lower.5ex\hbox{\gtsima}}
\def\lsim{\lower.5ex\hbox{\ltsima}}
\def\simgt{\lower.5ex\hbox{\gtsima}}
\def\simlt{\lower.5ex\hbox{\ltsima}}
\def\simpr{\lower.5ex\hbox{\prosima}}
\def\mean#1{\left< #1 \right>}
 
 \newcommand*\oline[1]{%
  \vbox{%
    \hrule height 0.5pt
    \kern0.25ex
    \hbox{%
      \kern-0.1em
      \ifmmode#1\else\ensuremath{#1}\fi
      \kern-0.1em
    }
  }
}

\begin{document}

\title[Detecting high-$z$ galaxies in the NIRB]{Detecting high-$z$ galaxies in the Near Infrared Background}
 \author[Yue et al.]{Bin Yue$^{1}$, Andrea Ferrara$^{1, 2}$,
 K\'ari Helgason$^{3}$  \\
 $^1$Scuola Normale Superiore, Piazza dei Cavalieri 7, I-56126 Pisa, Italy\\
 $^2$Kavli IPMU (WPI), Todai Institutes for Advanced Study, 
The University of Tokyo, 5-1-5 Kashiwanoha, Kashiwa 277-8583, Japan\\
 $^3$Max Planck Institute for Astrophysics, Karl-Schwarzschild-Str. 1, 85748 Garching, Germany\\
 }

\maketitle

\begin{abstract}
Emission from high-$z$ galaxies must unquestionably contribute to the near-infrared background (NIRB). However, this contribution has so far proven difficult to isolate even after subtracting the resolved galaxies to deep levels. Remaining NIRB fluctuations are dominated by unresolved low-$z$ galaxies on small angular scales, and by an unidentified component with unclear origin on large scales ($\approx 1000''$). In this paper, by analyzing mock maps generated from semi-numerical simulations and empirically determined $L_{\rm UV} - M_{\rm h}$ relations, we find that fluctuations associated with galaxies at $5 < z < 10$ amount to several percent of the unresolved NIRB flux fluctuations. We investigate the properties of this component for different survey areas and limiting magnitudes. In all cases, we show that this signal can be efficiently, and most easily at small angular scales, isolated by cross-correlating the source-subtracted NIRB with Lyman Break Galaxies (LBGs) detected in the same field by {\tt HST} surveys.  This result provides a fresh insight into the properties of reionization sources. 
\end{abstract}


\section{Introduction}
The cosmic infrared background (CIB or CIRB) contains a considerable fraction of the collective radiation emitted by stars through the cosmic time. Stars in the Epoch of Reionization (EoR) have the bulk of their radiation redshifted into the near-infrared band ($\sim$0.7 - 5 $\mu$m) and the CIB measured in this band is specifically named ``near-infrared background" (NIRB). 
As such, the NIRB offers
a unique opportunity to study faint high-$z$ galaxies that remain largely undetected in deep galaxy surveys (see e.g. \citealt{2006MNRAS.367L..11S, 2006ApJ...646..703F,2010ApJ...710.1089F,2012ApJ...750...20F,2013ApJ...764...56F,2013MNRAS.433.2047F}). This is particularly important as these objects are commonly believed to provide most of the ionizing power that drives cosmic reionization \citep{2007MNRAS.380L...6C,2011MNRAS.410..775R,2011MNRAS.414..847S}. Besides, the NIRB might also help characterizing the stellar populations of the first cosmic systems 
\citep{2003MNRAS.339..973S,2006MNRAS.368L...6S,2002MNRAS.336.1082S,2004ApJ...608....1K,2005Natur.438...45K,2003MNRAS.342L..25M,2004MNRAS.351L..71C,2004ApJ...606..611C}. 

Most recent studies have converged on the prediction that on scales of $\approx 1000''$ the fluctuation level from normal star-forming galaxies at $z \gsim 5$ is $\approx 10^{-3}$~nWm$^{-2}$sr$^{-1}$ at 3.6~$\mu$m \citep{2012ApJ...756...92C,2013MNRAS.431..383Y,2016MNRAS.455..282H}. However, extracting such signal from available data has been so far very challenging,  as even when the deepest galaxy
subtraction from NIRB maps is applied, the remaining flux fluctuations\footnote{NIRB studies usually concentrate on  fluctuations rather than absolute flux, as the latter is difficult to measure due to the presence of an overwhelming foreground. However, as the foreground is rather smooth on scales at which the NIRB is measured, the fluctuations analysis is more robust -- see e.g. \citealt{2007ApJ...657..669T,2011ApJ...742..124M,2012ApJ...753...63K,2012Natur.490..514C}.} still cannot be ascribed to the known high-$z$ galaxy population. On small angular scales, most of the signal arises from unresolved low-$z$ galaxies (see the first analysis by \citealt{2002ApJ...579L..53K}).
On larger scales the measured power  (see e.g. \citealt{2004ApJ...608....1K,2005Natur.438...45K,2007ApJ...654L...1K,2007ApJ...654L...5K,2012ApJ...753...63K,2005ApJ...626...31M, 2011ApJ...742..124M,2015arXiv150405681S,2007ApJ...659L..91C,2012Natur.490..514C}) is $\gsim 100$ times larger than the low-$z$ galaxies \citep{2012ApJ...752..113H}, and $\gsim 1000$ times larger than that expected from high-$z$ normal star-forming galaxies and first stars \citep{2012ApJ...756...92C,2013MNRAS.431..383Y}. Therefore it must be attributed to some, yet unknown, alternative sources. Basically, two different explanations have been proposed  for the origin of such large scale ($\sim1000''$) ``power excess". They involve either early accreting black holes \citep{2013MNRAS.433.1556Y,2014MNRAS.440.1263Y} which could explain the detected NIRB-cosmic X-ray background coherence \citep{2013ApJ...769...68C}, or ``intrahalo light'' that radiated by stars ejected from their parent galaxies during merger events \citep{2012Natur.490..514C,2014Sci...346..732Z}. 

At present it is unclear which hypothesis should be preferred. Besides, it is also possible that first stars or black holes are much more abundant than in the standard theoretical framework \--- for example if the slope of the density fluctuations power spectrum sightly deviates from the standard one, the number of small halos (thus stars or black holes therein) could be boosted exponentially \--- they can then provide sufficient radiation power. To be consistent with electron scattering reionization bounds, at the same time the ionizing photon escape probability must be rather low (for a detailed discussion see \citealt{2016MNRAS.455..282H}). To further complicate the interpretation, recent observations \citep{2014Sci...346..732Z} show that, on large scales and at least for the 1.1 and 1.6~$\mu$m bands, diffuse Galactic light (DGL) might provide a non-negligible flux contribution. However \citet{2007ApJ...654L...5K} (see \citealt{2010ApJS..186...10A} as well) found that at the 3.6 and 4.5 $\mu$m bands the DGL or the Galactic cirrus component is largely subdominant. 

For the above reasons, it is urgent to devise new strategies that put our understanding on firmer grounds. In order to isolate the signal from increasingly high redshifts, with sufficient depth and angular resolution one can in principle remove foreground galaxies down to extremely faint levels. This is challenging for current instrumentation, and also not an easy task for the {\tt JWST}, as the signal from high-$z$ galaxies is expected to be subdominant even at $\sim$32 AB mag \citep{2016MNRAS.455..282H}.  
Alternatively, cross-correlation studies seem promising. The NIRB-HI 21cm line cross-correlation \citep{2014MNRAS.440..298F,2014ApJ...790..148M} has the advantage that it selectively picks up the signal from reionization sources.  Also, the NIRB, if produced by sources in the EoR, would be cross-correlated with the CMB through the thermal Sunyaev-Zeldovich effect. This correlation could be seen in the forthcoming Euclid-based all-sky CIB maps \citep{2014ApJ...797L..26A}.  In addition to cross-correlation, a recent work \citep{2015ApJ...804...99K} proposed the use of Lyman-break tomography to constrain the NIRB contribution from sources above a certain redshift.

\citet{2007ApJ...666L...1K} analyzed the cross-correlation between the ACS-detected faint sources and the source-subtracted NIRB. They found significant correlation on small scales, implying that the faint ACS galaxies do contribute to NIRB fluctuations. However, these sources (or objects associated with them) cannot account for the large scale clustering as the measured correlation is negligible. Their analysis did not pay particular attention to the high-$z$ sources, therefore the information on high-$z$ galaxies cannot be directly derived from there.
It is surprising that so far little attention has been devoted to the search for high-$z$ normal star-forming galaxy signatures in the NIRB, given that deep galaxy surveys have made tremendous progresses in obtaining the UV luminosity functions (LF) up to $z=10$ \citep{2015ApJ...803...34B}, and the detection limits of Lyman break galaxy (LBG) surveys carried out by {\tt HST} have already reached $H\sim 29 - 31$ \citep{2013ApJS..209....6I,2011ApJ...737...90B}.  
Measuring their intensity fluctuations present (though tiny) in the NIRB is essentially one of the most exciting perspectives as it might convincingly show that NIRB fluctuation experiments can be used to study the first galaxies. 
Such an experiment is also complementary to more traditional galaxy surveys that derive the LF of the brightest galaxies among early populations. 

In summary, it is mandatory to show that the NIRB power spectrum signal of already known high-$z$ galaxies can be recovered robustly. Demonstrating a successful strategy will represent a major step forward in the methodology and allows to obtain, in addition to the clustering signal, other key quantities, as e.g. the colors of high-$z$ galaxies beyond the observed $H$ band. Colors, in turn, provide potentially information on stellar ages and initial mass function of the stars harbored by galaxies in the EoR.  

The idea we propose here is to isolate the targeted LBG signal and show the feasibility of statistically detecting reionization sources by cross-correlating deep LBG surveys with NIRB maps. To this aim, we: (a) construct large scale mock maps of the source-subtracted NIRB and LBG catalogs using semi-numerical simulations; (b) perform a cross-correlation analysis between the two data sets to extract the contribution of high-$z$ galaxies in the NIRB. The paper is organized as follows. In Sec. \ref{method} we describe the steps to construct the mock maps. In Sec. \ref{results} we present the analysis about the correlation coefficient and the colors. Conclusions and discussions are presented in Sec. \ref{conclusions}. Throughout this paper we use {\tt Planck} cosmological parameters: $\Omega_{\rm m} = 0.31, \Omega_\Lambda = 0.69, \Omega_{\rm b} = 0.048, n_s = 0.96, \sigma_8 = 0.82$ and $h = 0.68$ \citep{2014A&A...571A..16P}. All magnitudes are in the AB-system \citep{1983ApJ...266..713O}.

\section{Construction of mock maps}\label{method}

\subsection{ High-$z$ galaxies }
Using the code {\tt DexM}\footnote{\url{http://homepage.sns.it/mesinger/DexM___21cmFAST.html}}   \citep{2007ApJ...669..663M} we carry out semi-numerical simulations to get catalogs of halos with mass $M_{\rm h}\gsim 5\times10^8~M_\odot$ from $z = 5$ to 10 for every $\Delta z = 0.1$. We adopt a 400~Mpc box size that corresponds  to an angular size  of $\approx2.4$~deg at $z = 10$. We construct a cuboid by replicating the output boxes along the line-of-sight, adding random translations, rotations and reflections \citep{2005MNRAS.360..159B}. This is our light-cone since we assume that all lines-of-sight are parallel \--- an assumption that is convenient and safe enough when $z \gsim 5$.

To construct flux maps from the light-cone, we link galaxy luminosities to halo mass $M_{\rm h}$. The observed  LFs could be reproduced exactly if we derive the $L_{\rm UV} - M_{\rm h}$ relation through abundance matching, i.e. we force the number density of galaxies with luminosity $>L_{\rm UV}$ to match the number density of halos with  mass $>M_{\rm h}$. Formally, 
\begin{equation}
\int_{M_{\rm UV}} \Phi(M^\prime_{\rm UV},z) dM^\prime_{\rm UV} = \int_{M_h} \frac{dn}{dM^\prime_h}dM^\prime_h,
\end{equation}
where $\Phi$ is the UV LF at 1600~\AA~and we use the Schechter parameterization with the redshift-dependent fitting parameters given in \citet{2015ApJ...803...34B}. As a reference, in the derived $L_{\rm UV} - M_{\rm h}$ relations 
our minimum mass $5\times10^8~M_\odot$ corresponds  to an absolute magnitude $M_{\rm UV} = -10.5, -12.1, -13.0$ at $z=5, 8, 10$ respectively. Luminosity at other UV wavelengths is obtained through the luminosity-dependent Spectral Energy Distribution (SED) slope $\beta$ (i.e., $f_\lambda \propto \lambda^{\beta}$) in \citet{2014ApJ...793..115B}\footnote{$\beta=\beta_{-19.5}+\frac{d\beta}{dM_{\rm UV}}(M_{\rm UV}+19.5)$, the values of $\beta_{-19.5}$ and $\frac{d\beta}{dM_{\rm UV}}$ at $z= 4, 5, 6$ and 7 could be found in \citet{2014ApJ...793..115B}. For convenience of using this form at in-between redshifts, we fit $z$-dependent forms \citep{2015MNRAS.450.3829Y}: $\beta_{-19.5}=-1.97-0.06(z-6)$ and $\frac{d\beta}{dM_{\rm UV}}=-0.18-0.03(z-6)$. We use these fittings anyway when $5 < z < 10$.}. However, generally speaking this power-law only holds at $\lambda \lsim 2000 - 3000$~\AA, while we need luminosities at least until $4.5/(1+z)~\mu$m, say 7500~\AA~when $z=5$. Therefore at $\lambda>$2000~\AA~we use the SED template from {\tt Starburst99}\footnote{http://www.stsci.edu/science/starburst99/docs/default.htm} \citep{1999ApJS..123....3L,2005ApJ...621..695V,2010ApJS..189..309L}, adopting a continuous star formation mode, with metallicity 0.1~$Z_\odot$ and 200~Myr stellar age. The SB99 SEDs are normalized to match the power-law form at 2000~\AA.

\begin{figure*}
\centering{
\subfigure{\includegraphics[scale=0.4]{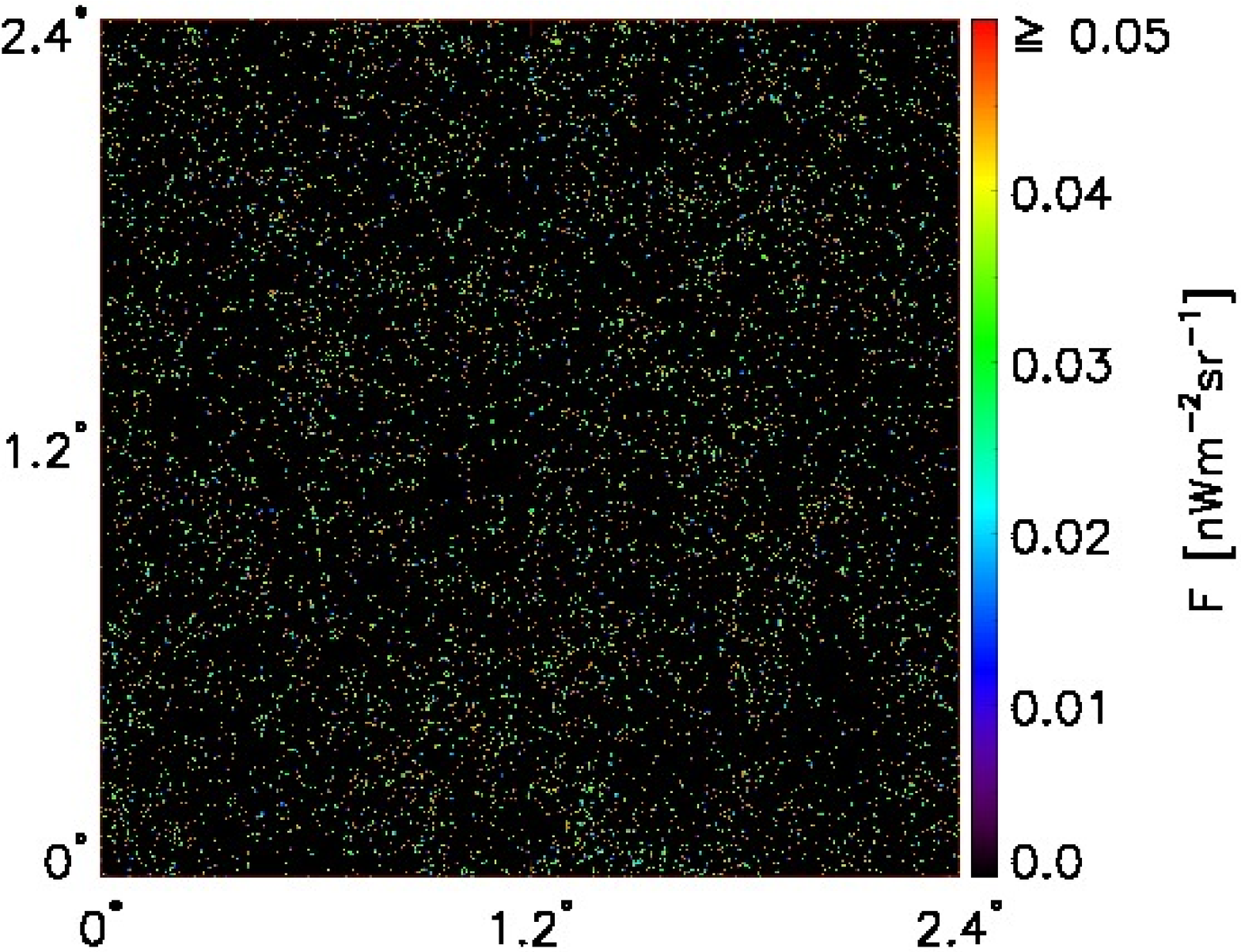}}
\subfigure{\includegraphics[scale=0.4]{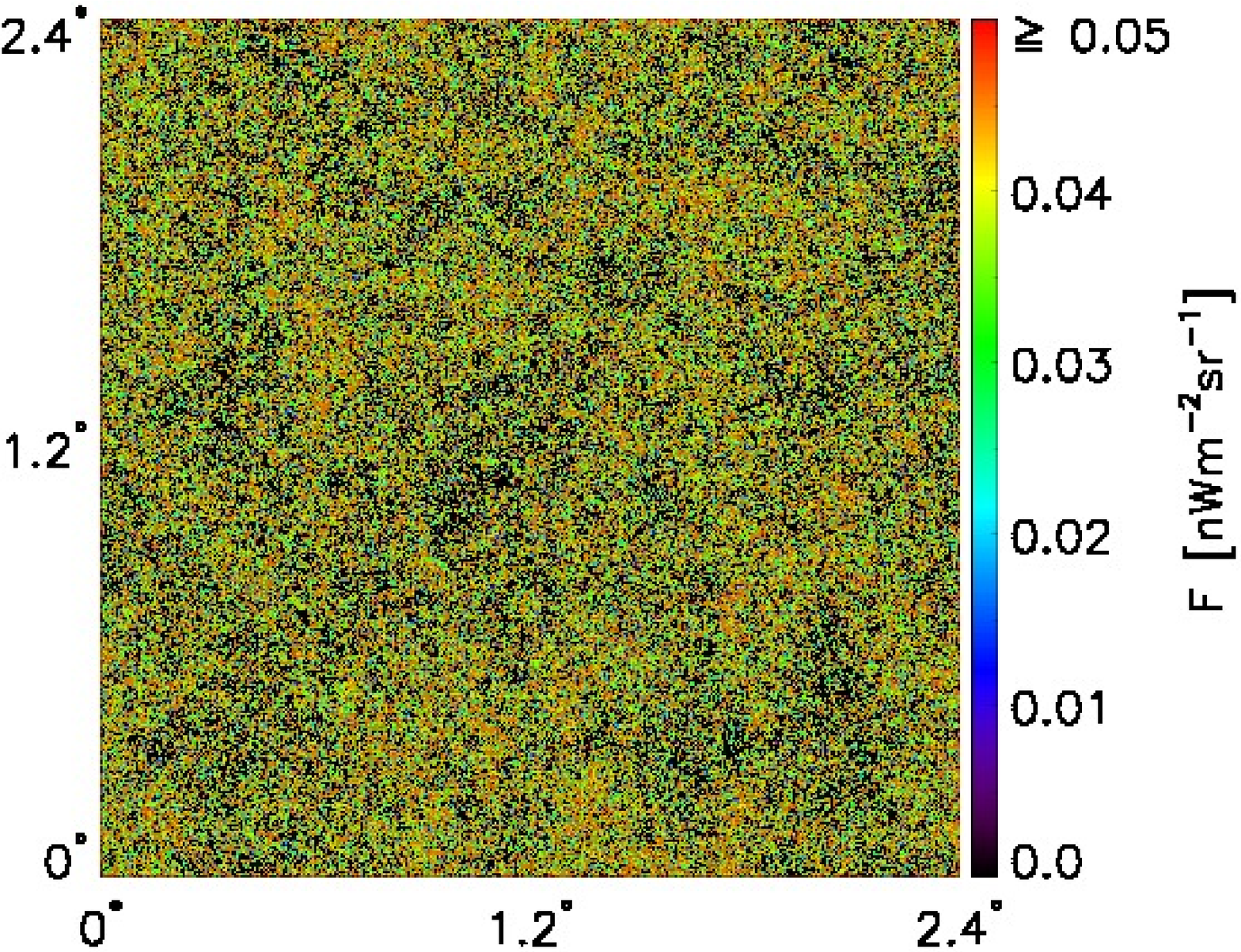}}
\subfigure{\includegraphics[scale=0.4]{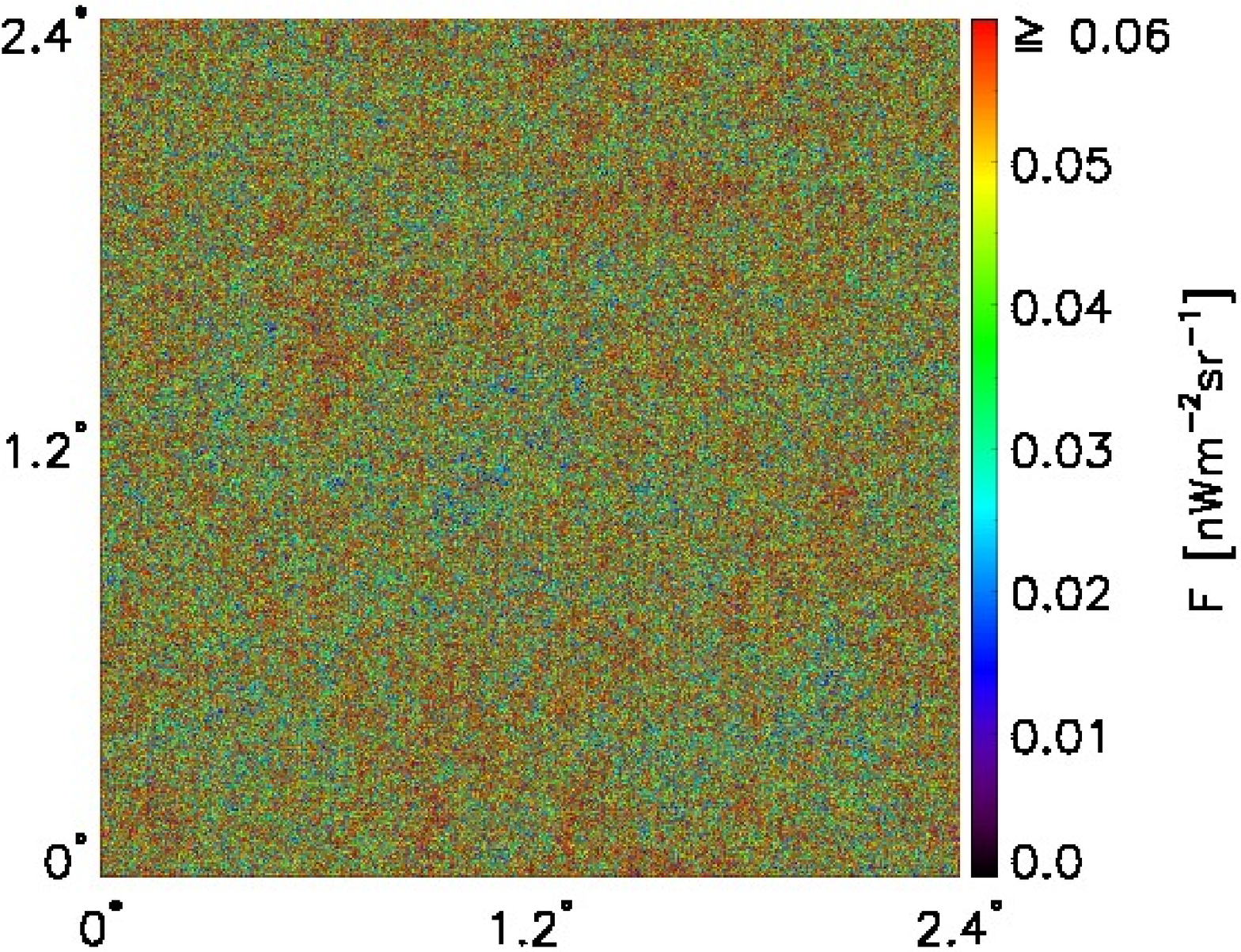}}
\subfigure{\includegraphics[scale=0.4]{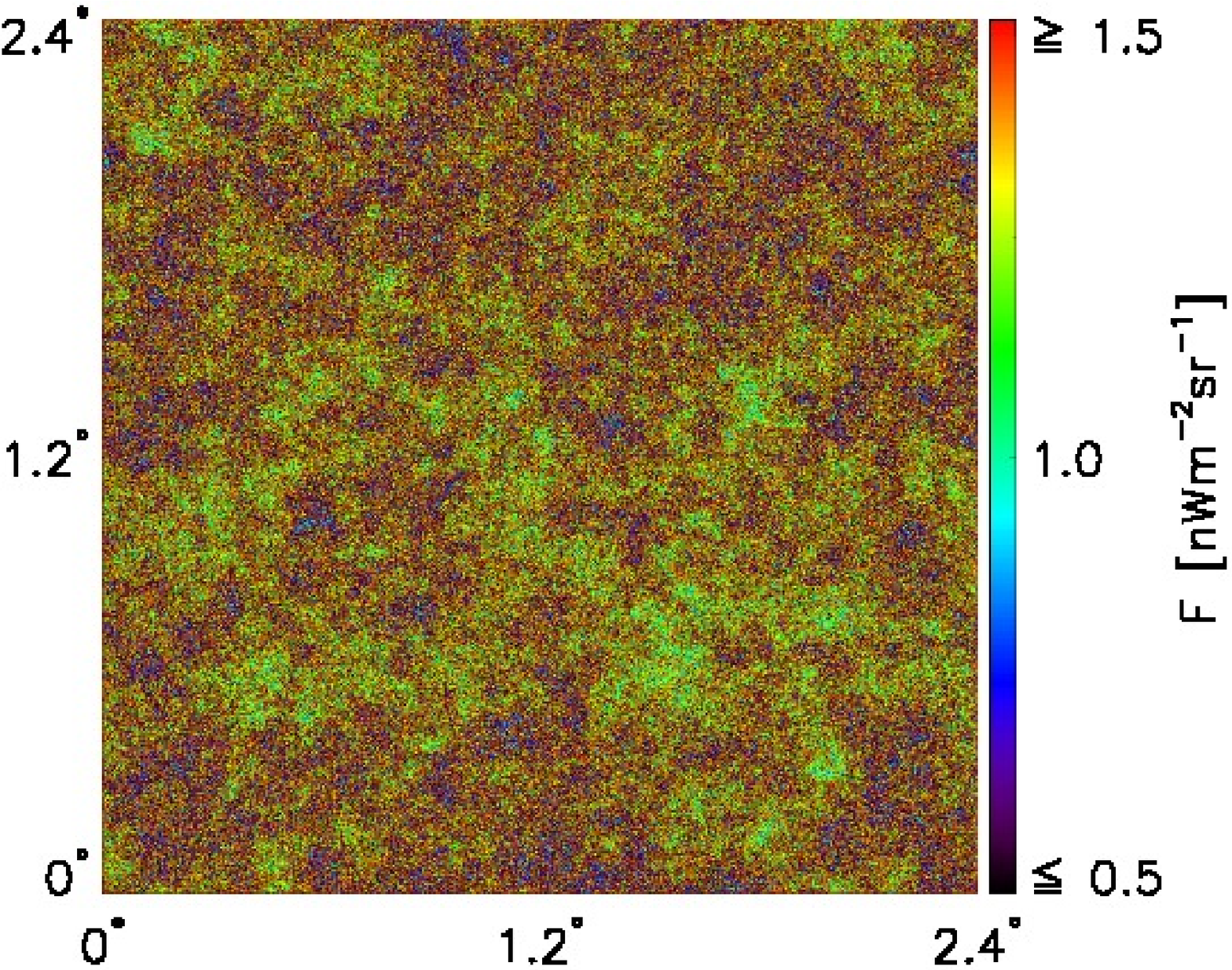}}
 \caption{{\it Upper:} The 1.6~$\mu$m flux map constructed from resolved LBGs with $H_{\rm lim} = 25$ (left) and $H_{\rm lim} = 27$ (right) respectively. {\it Lower:} Map of the 3.6~$\mu$m flux from galaxies with $5 < z < 10$ (left) and map of contamination at the same wavelength (right). The mean flux of the latter is 1 nWm$^{-2}$sr$^{-1}$.}
\label{maps}
}
\end{figure*}

The flux received in each pixel in the map is the sum of radiation from all galaxies seen by the pixel,
\begin{equation}
F(\nu_0)=\nu_0\frac{1}{(\theta_{\rm pix})^2}\sum_j \frac{L^j(\nu)(1+z_j)}{4\pi r_j^2(1+z_j)^2},
\end{equation}
where $\nu=\nu_0(1+z_j)$, $z_j$ is the redshift of the $j$-th halos\footnote{We do not model photometric redshift uncertainties, because the redshift range considered here $z = 5-10$ is much larger than the uncertainties.} in the solid angle $(\theta_{\rm pix})^2$ (we adopt $\theta_{\rm pix}=3.6''$ for all mock maps in this work),
 $r_j$ is its comoving distance.
In Fig. \ref{maps} we show the 3.6~$\mu$m flux map (bottom left panel) from all galaxies between $z = 5 -10$ 
(flux from galaxies at $z > 10$ is negligible compared with galaxies at lower redshift, we ignore it here).  
  
Finally, we construct the flux map of LBGs at $5<z<10$. To take into account selection effects, we assume a completeness function of the form
\begin{equation}
f(m)=0.5[1-{\rm erf}(m-m_{\rm lim})],
\end{equation}
where erf is the error function, and $m_{\rm lim}$ is the limiting magnitude. When constructing the flux map, for each LBG with apparent magnitude $m$, we generate an uniformly distributed number $x_r$. Flux from these galaxies is added to the map only if $x_r \le f(m)$.
In Fig. \ref{maps} we show the 1.6~$\mu$m\footnote{In this paper we only discuss the 1.6~$\mu$m flux maps of LBGs, because we consider a redshift range from $z = 5$ to 10. For shorter wavelengths all procedures (and conclusions) are similar, with the only exception of a slightly smaller signal due to the Lyman dropout of $z\gtrsim 8$ galaxies.}
flux map constructed from the LBGs for $H$-band limiting magnitudes $H_{\rm lim} = 25$ (top left) and $H_{\rm lim} = 27$ (top right) respectively.

\subsection{NIRB contamination maps}

\begin{figure}
\centering{
\includegraphics[scale=0.4]{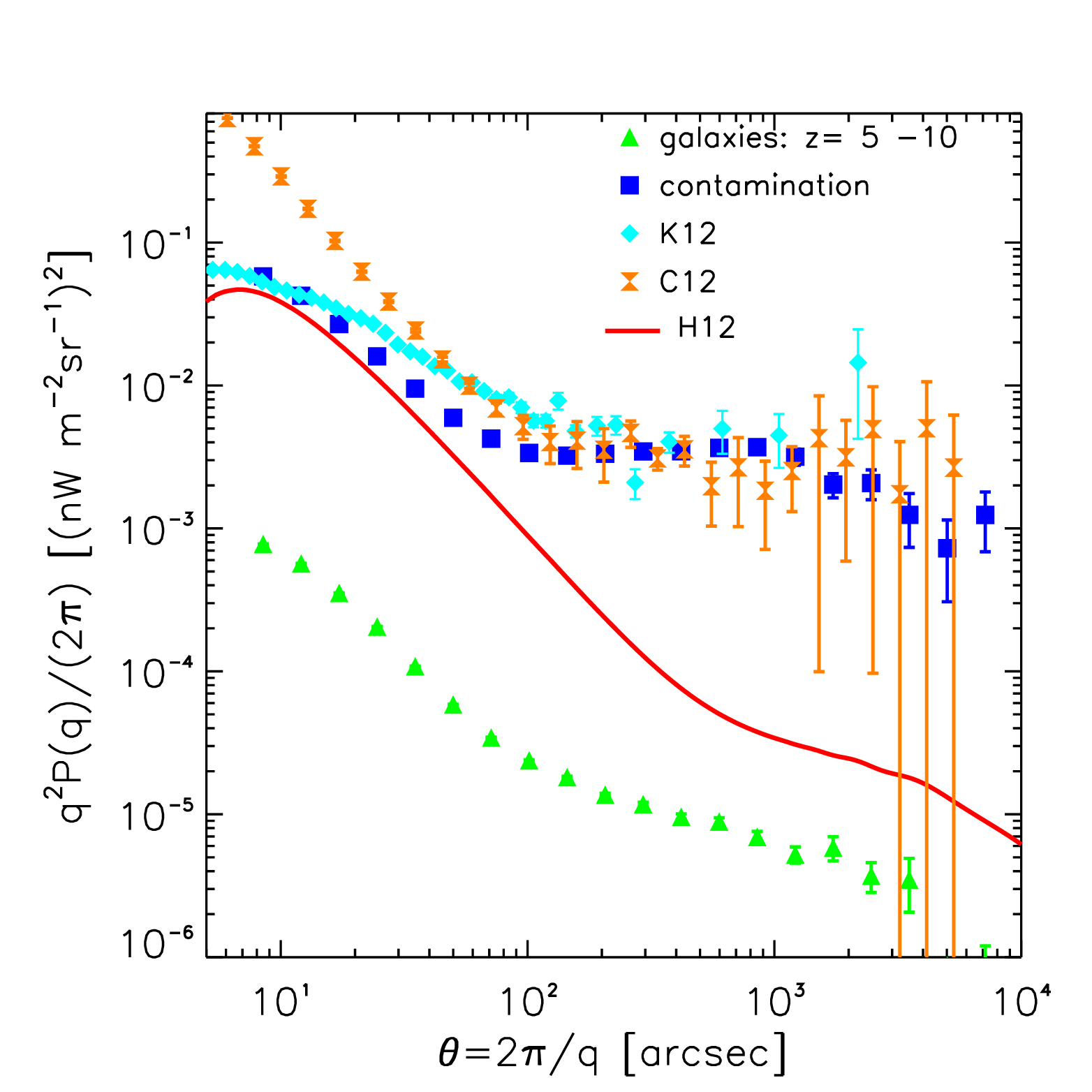}
 \caption{Angular power spectrum of the 3.6~$\mu$m flux map for $5 < z < 10$ galaxies (triangles) and contamination (squares). Errorbars show uncertainties due to limited number of Fourier modes in each $q$ bin. As a comparison, in the same panel we also plot  the \citet{2012ApJ...752..113H} model for $z < 5$ galaxies (solid line), and the observational points in \citet{2012ApJ...753...63K} (diamonds) and \citet{2012Natur.490..514C} (hourglasses).  
On large scales ($\theta \gsim 100''$) the angular power spectrum of our co-added map (quite similar to squares in the panel) is consistent with both observations. Compared with K12, at $\theta \lsim 100''$ our prediction slightly falls short, probably because the non-linear clustering of low-$z$ galaxies is not modeled here. The C12 observations have a shallower (i.e. $\sim24$ mag) source subtraction, hence a higher shot-noise level.
}
\label{APS}
}
\end{figure}

In addition to the flux from high-$z$ ($z > 5$) galaxies, the observed NIRB also contains radiation from unresolved, low-$z$ galaxies, and an excess radiation from unknown sources \citep{2013MNRAS.433.1556Y,2014MNRAS.440.1263Y,2012Natur.490..514C,2014Sci...346..732Z}. We collectively refer to these two components as {\it contamination}, since in this work the targeted signal is the flux from high-$z$ galaxies. To model such contaminating signal we construct random maps with mean flux 1.0 (0.7) nWm$^{-2}$sr$^{-1}$ at 3.6 (4.5) $\mu$m and reproduce the sum of (i) the angular power spectrum of the power {\it excess}  (see \citealt{2013MNRAS.433.1556Y}) matching available observations \citep{2012Natur.490..514C,2012ApJ...753...63K}, and (ii) the angular power spectrum of low-$z$ galaxies \citep{2012ApJ...752..113H} producing shot noise level matching 
\citealt{2012ApJ...753...63K} ($P_{\rm SN} = 4.8\times10^{-11}$nW$^{2}$m$^{-4}$sr$^{-1}$ at 3.6~$\mu$m and $P_{\rm SN} = 2.2\times10^{-11}$nW$^{2}$m$^{-4}$sr$^{-1}$ at 4.5~$\mu$m, the corresponding subtraction magnitude is $\sim25$ mag). 
The steps are:
\begin{itemize}
\item A white noise map, i.e. a Gaussian random field, is generated.
\item This map is then transformed into spatial-frequency space by FFT.
\item For each complex number in frequency space, its modulus is rescaled to be $\sqrt{P(q)}$, where $P$ is the given power spectrum and $q$ is the spatial-frequency. The zero-frequency ($q = 0$) element is set to be the mean flux.
\item The above map is then transformed back into real space by inverse FFT, resulting in a synthetic image with the same 2-point clustering properties as the measured $P(q)$.
\end{itemize}
In Fig. \ref{maps} we plot a single realization of the contamination map at 3.6~$\mu$m as an example (bottom right). The contamination is not correlated with the high-$z$ galaxy component; however, it adds noise to the cross-correlation signal. To account for the statistical variance of the contamination, we make 30 independent realizations. In Fig. \ref{APS} we show the angular power spectrum of bottom panels in Fig. \ref{maps}. All maps are convolved with a circular symmetric {\tt Spitzer} PSF before further analysis. 

\section{Cross-correlation of LBG\lowercase{s} and NIRB}\label{results}

\subsection{the correlation coefficient}
We first analyze the correlation coefficient between the LBG flux map and the NIRB map. It is defined as
 \begin{equation}
 {\cal R}=\frac{\mean{\delta F_{1.6} \delta F_{\lambda_0}}}{\sqrt{ \mean{\delta F^2_{1.6}} \mean{  \delta F^2_{\lambda_0}}    }  },
 \end{equation}
 where $\lambda_0$ refers to either 3.6 $\mu$m or 4.5 $\mu$m, $\delta F_{1.6}$ and $\delta F_{\lambda_0}$ are the flux fluctuations of the same pixel (zero lag) at those two wavelengths. The brackets refer to the pixel-averaged fluctuations. The correlation coefficient indicates the fraction of sources contributing to common signals.  Before calculating the correlation coefficient we smooth both the NIRB maps and the LBG flux maps 
by a real space top-hat window function with diameter $\theta=10''$. This is to suppress the instrumental noise since in the measured NIRB maps \citep{2012ApJ...753...63K} the instrumental noise is negligible at $\theta \gsim 10''$.
The smoothing is not related to the LBG detection limits since it is performed on maps constructed from LBGs already present in catalogs.
To mimic surveys with different areas, we cut out sub-maps with different areas from the full map. We choose three areas:
$(0.036)^2$, $(0.3)^2$ and $(1.2)^2$~deg$^2$, representing a survey region similar to HUDF/XDF, UDS, and an hypothetical larger field, respectively. 
 
\begin{figure}
\centering{
\includegraphics[scale=0.4]{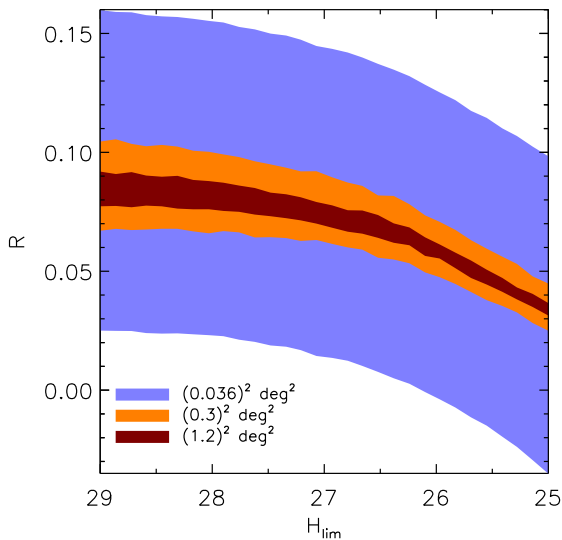}
\caption{
Correlation coefficient between the 3.6~$\mu$m source-subtracted NIRB and LBG flux maps vs. $H$-band limiting magnitudes for three different map areas. Filled regions bracket the 68.3\% probability ranges around the peaks.
All fields are smoothed on scale $\theta=10''$. }	
\label{correlation_coefficient}
}
\end{figure}

Before calculating the correlation coefficient, in both maps we mask pixels containing galaxies brighter than mag $\sim25$ at {\it either} 3.6~$\mu$m or 4.5~$\mu$m. From this procedure we obtain the source-subtracted NIRB map. The correlation coefficient vs. LBG limiting magnitude is shown in Fig. \ref{correlation_coefficient}. The filled regions  contain 68.3\% probability of all sub-map samples. 
Note that we have 30 random realizations for each full map ((2.4)$^2$ deg$^2$), so even for the $(1.2)^2$~deg$^2$ case there are 120 samples, sufficient to compute the correlation coefficient variance.  

Fig. \ref{correlation_coefficient} shows that it is indeed feasible to detect the cross-correlation signal from the mock maps, even for a relatively shallow survey with $H_{\rm lim}\sim25$, for which ${\cal R}\approx 0.04$. By pushing the limiting magnitude to fainter values, the correlation coefficient rapidly increases by a factor $\sim2$, and then slowly approaches $\approx 0.09$ at $H_{\rm lim}\sim29$. It is worth noting that in small fields the measured correlation coefficient has a $\sim 30 - 80\%$ relative field-by-field scatter for $H_{\rm lim} \gsim 26$, and even larger scatter at $H_{\rm lim} < 26$. In some extreme cases there would be no or negative cross-correlation detected, due to the small number of LBGs contained in the fields.

\begin{figure}
\centering{
\includegraphics[scale=0.4]{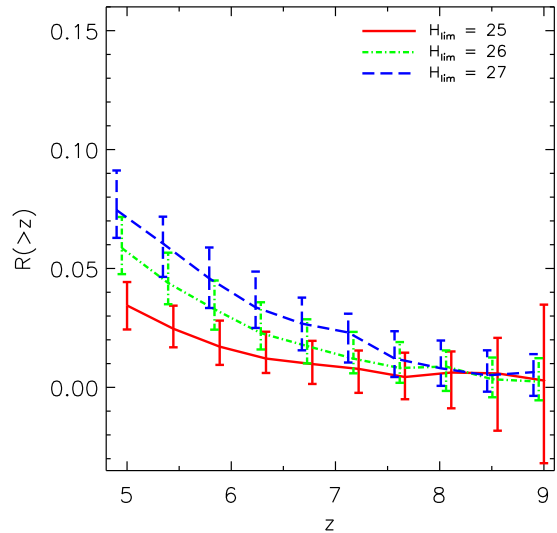}
\caption{ 
Contribution to correlation coefficient from LBGs with  $>z$. The 68.3\% probability ranges are plotted by errorbars. For displaying purpose we slightly shift the x-positions of some curves.}	
\label{correlation_coefficient_z}
}
\end{figure}

To elucidate the differential contribution of LBGs, for a $(0.3)^2$ deg$^2$ field, we further show in Fig. \ref{correlation_coefficient_z} the correlation coefficient from LBGs at $>z$ for $H_{\rm lim} = 25, 26$ and 27, respectively. For example, LBGs at $z>8$ contribute ${\cal R}\approx 0.01$ at $H\lsim 27$.

Our results are for NIRB maps with a fixed shot noise level matching \citet{2012ApJ...753...63K} measurements.  Since the shot noise is mainly from low-$z$ galaxies, an increased subtraction depth implies a relatively larger number of low-$z$ galaxies (with respect to the high-$z$ ones) are resolved and removed. As a result we expect a higher correlation coefficient. On the other hand, if we resolve more LBGs, the cross-power signal also becomes stronger. Thus, the strength of the cross-power signal vs. different LBGs detection depths can be used to determine the differential contribution of early galaxies.  
 
\subsection{Detectability vs. scales}

\begin{figure}
\centering{
\includegraphics[scale=0.4]{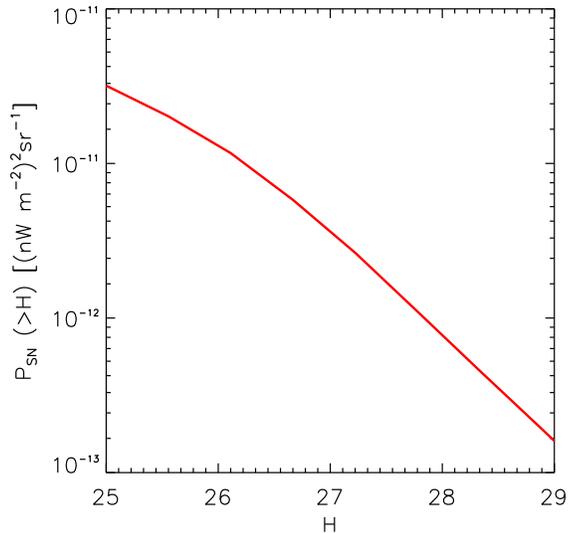}
\caption{Shot noise due to LBGs with magnitude $>H$ and at $5 < z < 10$.}	
\label{PSN}
}
\end{figure}

So far we have investigated the behavior of the correlation coefficient by at a specific smoothing scale $\theta=10''$. We now examine its dependence on angular scale. To this aim we re-define the correlation coefficient in the spatial-frequency domain via the power spectrum
\begin{equation} \label{eqn_coherence}
{\cal R}_q(\theta) = \frac{P_{\rm IR\times G}(\theta)}{  \sqrt{P_{\rm IR}(\theta)\times P_{\rm G}(\theta)} },
\end{equation}
where $P_{\rm 1\times 2}(\theta)$ is the cross-power spectrum and $P_{\rm 1}$,$P_{\rm 2}$ are the auto-power spectra. They are all calculated using the 2D FFT. Since $\delta F^2 \simeq q^2P/2\pi$, ${\cal R}_q \sim {\cal R}$ for the same $\theta$. 
First we derive the shot noise due to LBGs fainter than a given H-magnitude (Fig. \ref{PSN}). It is the value of $P_{\rm G}$ taken on a sufficiently small scale $\theta \approx 14''$ (a stability check of the results for scales  up to $\approx40''$ has been performed). As we will point out in the following,  shot noise  is the most important term contributing to the  cross-correlation. The shot noise level shown here is a useful reference for possible follow-up work on real data.  
 
\begin{figure}
\centering{
\includegraphics[scale=0.4]{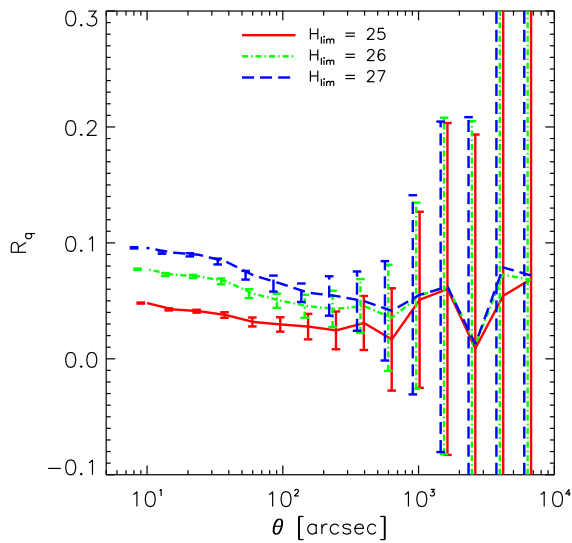}
\caption{The ${\cal R}_q$ between the source-subtracted NIRB and the flux maps constructed from LBGs with different $H_{\rm lim}$. The errorbars here show the r.m.s of 30 random realizations. For displaying purpose we slightly shift the x-positions of $H_{\rm lim} = 26, 27$ curves.}
\label{correlation_coefficient_q}
}
\end{figure}

The ${\cal R}_q(\theta)$ between the source-subtracted NIRB at $3.6~\mu$m and the 1.6~$\mu$m flux map of LBGs with $H_{\rm lim} =  25, 26$ and 27 is shown by Fig. \ref{correlation_coefficient_q}. The errorbars are the r.m.s of 30 samples each one using  different random contamination realizations. 

The cross-power spectrum includes both the shot noise and clustering terms. Shot noise is from sources common to both NIRB and LBG flux, and it is dominant on scales $ \theta \lsim 100''$. On larger scales, the clustering term progressively takes over. This term arises from all sources sharing the same large scale structure, including galaxies even fainter than the LBG limiting magnitude. As a consequence, in principle the clustering term could allow the detection of fainter galaxies in the source-subtracted NIRB through the cross-correlation with relatively bright LBGs. However, as shown by Fig. \ref{correlation_coefficient_q} the cross-correlation is more easily detected on small scales. Although the clustering term may contain more information, the main difficulty to be overcome in order to efficiently use this strategy is that even for a relatively large survey area of  $(2.4)^2$~deg$^2$ (i.e. our full map), the signal-to-noise ratio never exceeds $\sim 3$ at $\theta \gsim300''$ for $H_{\rm lim} < 27$.  While increasing the limiting magnitude to $>27$ does not help much to this aim, the noise could be reduced by using larger survey areas, as expected with, e.g. EUCLID and WFIRST.

\subsection{Colors}

\begin{figure}
\centering{
\includegraphics[scale=0.4]{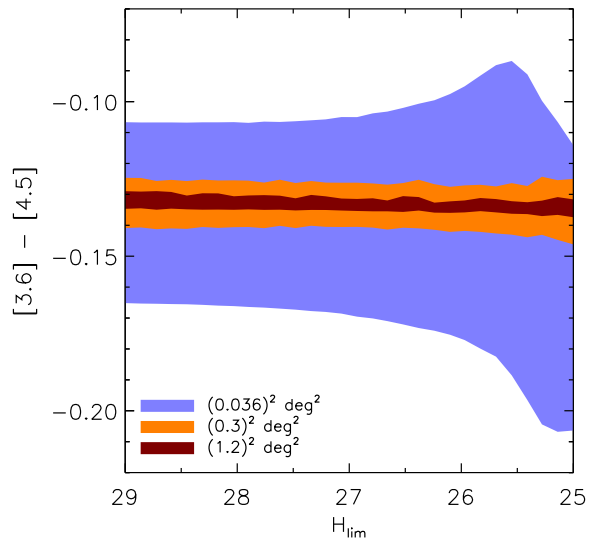}
\caption{The magnitude difference, $[3.6]-[4.5]$, of fluctuations due to unresolved galaxies in the source-subtracted NIRB maps as a function of the survey limiting magnitude. Filled regions indicate 68.3\% probability ranges.}	
\label{color}
}
\end{figure}

From our mock maps it is also possible to characterize the colors, i.e. the ratio between flux in two bands, of unresolved galaxies in NIRB observations. With the assumption 
\begin{equation}
\frac{F_{4.5}}{F_{3.6}}\approx \frac{\mean{ (\delta F_{4.5} \delta F_{1.6})_\theta  }} { \mean{ (\delta F_{3.6} \delta F_{1.6})_\theta }  },
\end{equation} 
we derive the magnitude difference as 
\begin{align}
[3.6]-[4.5] &= 2.5{\rm log}\left( \frac{4.5F_{4.5}}{3.6F_{3.6}}  \right) \nonumber \\
&\approx2.5{\rm log}\left( \frac{4.5}{3.6}  \frac{\mean{ (\delta F_{4.5} \delta F_{1.6} )_\theta }} { \mean{ (\delta F_{3.6} \delta F_{1.6} )_\theta}  }   \right)  
\end{align} 
Such magnitude difference, which might considerably vary for individual galaxies, represents the combined and weighted color of the unresolved galaxy population in these bands. In practice the PSF adds a bias to the measured flux ratio. Therefore before calculating the magnitude difference it would be necessary to first deconvolve the PSF from each map. We skip this step here and directly calculate the magnitude difference on maps without PSF convolution. Again we specify $\theta=10''$. The predicted color as a function of $H_{\rm lim}$ is reported in Fig. \ref{color}, allowing us to conclude that the magnitude difference which is $\sim$-0.13 mag, could be detected by cross-correlating survey maps with area $\gsim (0.3)^2$ deg$^2$. The figure reiterates that smaller area fields would be affected by bias effects.

 \section{Conclusions}\label{conclusions} 

Motivated by the fact that LBG surveys carried out by the {\tt HST} have reached detection limits much deeper than the NIRB measured by {\tt Spitzer} at longer wavelengths, in this paper we investigated the feasibility to pick the resolved LBGs component out of the NIRB by a cross-correlation analysis. Our investigations were based on mock maps constructed from semi-numerical simulations of halo formation, and empirically determined $L_{\rm UV} - M_{\rm h}$ relations.
 
We found that in the NIRB observed at 3.6 and 4.5~$\mu$m and with sources subtracted down to $\sim$25 apparent magnitude, at smallest scales where the shot noise dominates, about 10\%  of the flux fluctuations arises from LBGs in $5 < z < 10$ and with $H \lsim 29$. Such faint galaxies have already been resolved in the existing deep surveys.  However, this fractional contribution, if measured from narrow fields with area $\sim (0.036)^2$~deg$^2$ (as the HUDF/XDF), could vary from $\sim3\%$ to $\sim16\%$. 
If the limiting magnitude is decreased to $H \sim 27$, the fractional contribution decreases to about 8\%. If in this case we consider a larger field, e.g. with area similar to the UDS ($\sim (0.3)^2$~deg$^2$), then the correlation coefficient  varies in the narrower range $\sim 6\%-9\%$. 
We remind that the variance at hand here is due to both the large-scale inhomogeneity of the signal and the contamination: it is the field-to-field variance of the correlation itself. We do not model errors introduced by data analysis procedures, as for example mask effects. However, at least theoretically we have shown that the contribution from the faintest galaxies could be isolated from the NIRB through the cross-correlation analysis. 

We pointed out that it is still challenging to use the cross-correlation arising from the clustering term of the cross-power spectrum to study galaxies unresolved not only in NIRB observations, but also in LBG surveys. This term is dominant at $\gsim 200''$, but even for survey areas as large as $(2.4)^2$~deg$^2$, the signal  has a small significance, with a S/N ratio $\lsim3$.   Although very unlikely, if the NIRB clustering excess originates from LBGs we would detect a correlation coefficient close to 1 at scales where the clustering term dominates. Stated differently, a measured correlation coefficient $ {\cal R} \ll 1$ would actually rule out this scenario.

Our results have interesting implications also for the colors of high-$z$ galaxy population. The colors of LBGs detected at wavelengths $< 1.6~\mu$m but too faint to be individually resolved by existing telescopes at  3.6 and 4.5~$\mu$m, can be obtained by using the cross-correlation with the NIRB in these two longer wavelength bands.

It is worth noting that the predictions presented in this paper assume a contaminating signal in the form of NIRB fluctuation excess which originates  from direct collapse black holes at high $z$ \citep{2013MNRAS.433.1556Y}. If however the excess is found to arise more locally, we might be able to model or subtract it more accurately, thereby making the signal calculated in this paper more easily detectable, i.e. ${\mean {\cal R}} \gsim 0.1$. 
 
Importantly, the study proposed here can be further developed to infer the properties of even fainter high-$z$ galaxy populations that are currently inaccessible to direct telescopic detection. This could, for example, be accomplished by a Lyman-break tomography study designed to isolate high-$z$ populations via multi-band cross-correlations without requiring prior LBG detections \citep{2015ApJ...804...99K}. 

\section*{ACKNOWLEDGMENTS}

We thank A. Kashlinsky for valuable comments on the manuscript.


\begin{thebibliography}{53}
\expandafter\ifx\csname natexlab\endcsname\relax\def\natexlab#1{#1}\fi

\bibitem[{{Arendt} {et~al}\mbox{.}(2010){Arendt}, {Kashlinsky}, {Moseley}, \&
  {Mather}}]{2010ApJS..186...10A}
{Arendt} R.~G., {Kashlinsky} A., {Moseley} S.~H., {Mather} J., 2010, \apjs,
  186, 10

\bibitem[{{Atrio-Barandela} \& {Kashlinsky}(2014)}]{2014ApJ...797L..26A}
{Atrio-Barandela} F., {Kashlinsky} A., 2014, \apjl, 797, L26

\bibitem[{{Blaizot} {et~al}\mbox{.}(2005){Blaizot}, {Wadadekar}, {Guiderdoni},
  {Colombi}, {Bertin}, {Bouchet}, {Devriendt}, \&
  {Hatton}}]{2005MNRAS.360..159B}
{Blaizot} J., {Wadadekar} Y., {Guiderdoni} B., {Colombi} S.~T., {Bertin} E.,
  {Bouchet} F.~R., {Devriendt} J.~E.~G., {Hatton} S., 2005, \mnras, 360, 159

\bibitem[{{Bouwens} {et~al}\mbox{.}(2011){Bouwens}, {Illingworth}, {Oesch},
  {Labb{\'e}}, {Trenti}, {van Dokkum}, {Franx}, {Stiavelli}, {Carollo},
  {Magee}, \& {Gonzalez}}]{2011ApJ...737...90B}
{Bouwens} R.~J. {et~al.}, 2011, \apj, 737, 90

\bibitem[{{Bouwens} {et~al}\mbox{.}(2014){Bouwens}, {Illingworth}, {Oesch},
  {Labb{\'e}}, {van Dokkum}, {Trenti}, {Franx}, {Smit}, {Gonzalez}, \&
  {Magee}}]{2014ApJ...793..115B}
{Bouwens} R.~J. {et~al.}, 2014, \apj, 793, 115

\bibitem[{{Bouwens} {et~al}\mbox{.}(2015){Bouwens}, {Illingworth}, {Oesch},
  {Trenti}, {Labb{\'e}}, {Bradley}, {Carollo}, {van Dokkum}, {Gonzalez},
  {Holwerda}, {Franx}, {Spitler}, {Smit}, \& {Magee}}]{2015ApJ...803...34B}
{Bouwens} R.~J. {et~al.}, 2015, \apj, 803, 34

\bibitem[{{Cappelluti} {et~al}\mbox{.}(2013){Cappelluti}, {Kashlinsky},
  {Arendt}, {Comastri}, {Fazio}, {Finoguenov}, {Hasinger}, {Mather}, {Miyaji},
  \& {Moseley}}]{2013ApJ...769...68C}
{Cappelluti} N. {et~al.}, 2013, \apj, 769, 68

\bibitem[{{Choudhury} \& {Ferrara}(2007)}]{2007MNRAS.380L...6C}
{Choudhury} T.~R., {Ferrara} A., 2007, \mnras, 380, L6

\bibitem[{{Cooray} {et~al}\mbox{.}(2004){Cooray}, {Bock}, {Keatin}, {Lange}, \&
  {Matsumoto}}]{2004ApJ...606..611C}
{Cooray} A., {Bock} J.~J., {Keatin} B., {Lange} A.~E., {Matsumoto} T., 2004,
  \apj, 606, 611

\bibitem[{{Cooray} {et~al}\mbox{.}(2012{\natexlab{a}}){Cooray}, {Gong},
  {Smidt}, \& {Santos}}]{2012ApJ...756...92C}
{Cooray} A., {Gong} Y., {Smidt} J., {Santos} M.~G., 2012{\natexlab{a}}, \apj,
  756, 92

\bibitem[{{Cooray} {et~al}\mbox{.}(2012{\natexlab{b}}){Cooray}, {Smidt}, {de
  Bernardis}, {Gong}, {Stern}, {Ashby}, {Eisenhardt}, {Frazer}, {Gonzalez},
  {Kochanek}, {Koz{\l}owski}, \& {Wright}}]{2012Natur.490..514C}
{Cooray} A. {et~al.}, 2012{\natexlab{b}}, \nat, 490, 514

\bibitem[{{Cooray} {et~al}\mbox{.}(2007){Cooray}, {Sullivan}, {Chary}, {Bock},
  {Dickinson}, {Ferguson}, {Keating}, {Lange}, \&
  {Wright}}]{2007ApJ...659L..91C}
{Cooray} A. {et~al.}, 2007, \apjl, 659, L91

\bibitem[{{Cooray} \& {Yoshida}(2004)}]{2004MNRAS.351L..71C}
{Cooray} A., {Yoshida} N., 2004, \mnras, 351, L71

\bibitem[{{Fernandez} {et~al}\mbox{.}(2013){Fernandez}, {Dole}, \&
  {Iliev}}]{2013ApJ...764...56F}
{Fernandez} E.~R., {Dole} H., {Iliev} I.~T., 2013, \apj, 764, 56

\bibitem[{{Fernandez} {et~al}\mbox{.}(2012){Fernandez}, {Iliev}, {Komatsu}, \&
  {Shapiro}}]{2012ApJ...750...20F}
{Fernandez} E.~R., {Iliev} I.~T., {Komatsu} E., {Shapiro} P.~R., 2012, \apj,
  750, 20

\bibitem[{{Fernandez} \& {Komatsu}(2006)}]{2006ApJ...646..703F}
{Fernandez} E.~R., {Komatsu} E., 2006, \apj, 646, 703

\bibitem[{{Fernandez} {et~al}\mbox{.}(2010){Fernandez}, {Komatsu}, {Iliev}, \&
  {Shapiro}}]{2010ApJ...710.1089F}
{Fernandez} E.~R., {Komatsu} E., {Iliev} I.~T., {Shapiro} P.~R., 2010, \apj,
  710, 1089

\bibitem[{{Fernandez} \& {Zaroubi}(2013)}]{2013MNRAS.433.2047F}
{Fernandez} E.~R., {Zaroubi} S., 2013, \mnras, 433, 2047

\bibitem[{{Fernandez} {et~al}\mbox{.}(2014){Fernandez}, {Zaroubi}, {Iliev},
  {Mellema}, \& {Jeli{\'c}}}]{2014MNRAS.440..298F}
{Fernandez} E.~R., {Zaroubi} S., {Iliev} I.~T., {Mellema} G., {Jeli{\'c}} V.,
  2014, \mnras, 440, 298

\bibitem[{{Helgason} {et~al}\mbox{.}(2012){Helgason}, {Ricotti}, \&
  {Kashlinsky}}]{2012ApJ...752..113H}
{Helgason} K., {Ricotti} M., {Kashlinsky} A., 2012, \apj, 752, 113

\bibitem[{{Helgason} {et~al}\mbox{.}(2016){Helgason}, {Ricotti}, {Kashlinsky},
  \& {Bromm}}]{2016MNRAS.455..282H}
{Helgason} K., {Ricotti} M., {Kashlinsky} A., {Bromm} V., 2016, \mnras, 455,
  282

\bibitem[{{Illingworth} {et~al}\mbox{.}(2013){Illingworth}, {Magee}, {Oesch},
  {Bouwens}, {Labb{\'e}}, {Stiavelli}, {van Dokkum}, {Franx}, {Trenti},
  {Carollo}, \& {Gonzalez}}]{2013ApJS..209....6I}
{Illingworth} G.~D. {et~al.}, 2013, \apjs, 209, 6

\bibitem[{{Kashlinsky} {et~al}\mbox{.}(2004){Kashlinsky}, {Arendt}, {Gardner},
  {Mather}, \& {Moseley}}]{2004ApJ...608....1K}
{Kashlinsky} A., {Arendt} R., {Gardner} J.~P., {Mather} J.~C., {Moseley} S.~H.,
  2004, \apj, 608, 1

\bibitem[{{Kashlinsky} {et~al}\mbox{.}(2012){Kashlinsky}, {Arendt}, {Ashby},
  {Fazio}, {Mather}, \& {Moseley}}]{2012ApJ...753...63K}
{Kashlinsky} A., {Arendt} R.~G., {Ashby} M.~L.~N., {Fazio} G.~G., {Mather} J.,
  {Moseley} S.~H., 2012, \apj, 753, 63

\bibitem[{{Kashlinsky} {et~al}\mbox{.}(2005){Kashlinsky}, {Arendt}, {Mather},
  \& {Moseley}}]{2005Natur.438...45K}
{Kashlinsky} A., {Arendt} R.~G., {Mather} J., {Moseley} S.~H., 2005, \nat, 438,
  45

\bibitem[{{Kashlinsky} {et~al}\mbox{.}(2007{\natexlab{a}}){Kashlinsky},
  {Arendt}, {Mather}, \& {Moseley}}]{2007ApJ...666L...1K}
{Kashlinsky} A., {Arendt} R.~G., {Mather} J., {Moseley} S.~H.,
  2007{\natexlab{a}}, \apjl, 666, L1

\bibitem[{{Kashlinsky} {et~al}\mbox{.}(2007{\natexlab{b}}){Kashlinsky},
  {Arendt}, {Mather}, \& {Moseley}}]{2007ApJ...654L...5K}
{Kashlinsky} A., {Arendt} R.~G., {Mather} J., {Moseley} S.~H.,
  2007{\natexlab{b}}, \apjl, 654, L5

\bibitem[{{Kashlinsky} {et~al}\mbox{.}(2007{\natexlab{c}}){Kashlinsky},
  {Arendt}, {Mather}, \& {Moseley}}]{2007ApJ...654L...1K}
{Kashlinsky} A., {Arendt} R.~G., {Mather} J., {Moseley} S.~H.,
  2007{\natexlab{c}}, \apjl, 654, L1

\bibitem[{{Kashlinsky} {et~al}\mbox{.}(2015){Kashlinsky}, {Mather}, {Helgason},
  {Arendt}, {Bromm}, \& {Moseley}}]{2015ApJ...804...99K}
{Kashlinsky} A., {Mather} J.~C., {Helgason} K., {Arendt} R.~G., {Bromm} V.,
  {Moseley} S.~H., 2015, \apj, 804, 99

\bibitem[{{Kashlinsky} {et~al}\mbox{.}(2002){Kashlinsky}, {Odenwald}, {Mather},
  {Skrutskie}, \& {Cutri}}]{2002ApJ...579L..53K}
{Kashlinsky} A., {Odenwald} S., {Mather} J., {Skrutskie} M.~F., {Cutri} R.~M.,
  2002, \apjl, 579, L53

\bibitem[{{Leitherer} {et~al}\mbox{.}(2010){Leitherer}, {Ortiz Ot{\'a}lvaro},
  {Bresolin}, {Kudritzki}, {Lo Faro}, {Pauldrach}, {Pettini}, \&
  {Rix}}]{2010ApJS..189..309L}
{Leitherer} C., {Ortiz Ot{\'a}lvaro} P.~A., {Bresolin} F., {Kudritzki} R.-P.,
  {Lo Faro} B., {Pauldrach} A.~W.~A., {Pettini} M., {Rix} S.~A., 2010, \apjs,
  189, 309

\bibitem[{{Leitherer} {et~al}\mbox{.}(1999){Leitherer}, {Schaerer}, {Goldader},
  {Delgado}, {Robert}, {Kune}, {de Mello}, {Devost}, \&
  {Heckman}}]{1999ApJS..123....3L}
{Leitherer} C. {et~al.}, 1999, \apjs, 123, 3

\bibitem[{{Magliocchetti} {et~al}\mbox{.}(2003){Magliocchetti}, {Salvaterra},
  \& {Ferrara}}]{2003MNRAS.342L..25M}
{Magliocchetti} M., {Salvaterra} R., {Ferrara} A., 2003, \mnras, 342, L25

\bibitem[{{Mao}(2014)}]{2014ApJ...790..148M}
{Mao} X.-C., 2014, \apj, 790, 148

\bibitem[{{Matsumoto} {et~al}\mbox{.}(2005){Matsumoto}, {Matsuura}, {Murakami},
  {Tanaka}, {Freund}, {Lim}, {Cohen}, {Kawada}, \&
  {Noda}}]{2005ApJ...626...31M}
{Matsumoto} T. {et~al.}, 2005, \apj, 626, 31

\bibitem[{{Matsumoto} {et~al}\mbox{.}(2011){Matsumoto}, {Seo}, {Jeong}, {Lee},
  {Matsuura}, {Matsuhara}, {Oyabu}, {Pyo}, \& {Wada}}]{2011ApJ...742..124M}
{Matsumoto} T. {et~al.}, 2011, \apj, 742, 124

\bibitem[{{Mesinger} \& {Furlanetto}(2007)}]{2007ApJ...669..663M}
{Mesinger} A., {Furlanetto} S., 2007, \apj, 669, 663

\bibitem[{{Oke} \& {Gunn}(1983)}]{1983ApJ...266..713O}
{Oke} J.~B., {Gunn} J.~E., 1983, \apj, 266, 713

\bibitem[{{Planck Collaboration} {et~al}\mbox{.}(2014){Planck Collaboration},
  {Ade}, {Aghanim}, {Armitage-Caplan}, {Arnaud}, {Ashdown}, {Atrio-Barandela},
  {Aumont}, {Baccigalupi}, {Banday}, \& et~al.}]{2014A&A...571A..16P}
{Planck Collaboration} {et~al.}, 2014, \aap, 571, A16

\bibitem[{{Rai{\v c}evi{\'c}} {et~al}\mbox{.}(2011){Rai{\v c}evi{\'c}},
  {Theuns}, \& {Lacey}}]{2011MNRAS.410..775R}
{Rai{\v c}evi{\'c}} M., {Theuns} T., {Lacey} C., 2011, \mnras, 410, 775

\bibitem[{{Salvaterra} \& {Ferrara}(2003)}]{2003MNRAS.339..973S}
{Salvaterra} R., {Ferrara} A., 2003, \mnras, 339, 973

\bibitem[{{Salvaterra} \& {Ferrara}(2006)}]{2006MNRAS.367L..11S}
{Salvaterra} R., {Ferrara} A., 2006, \mnras, 367, L11

\bibitem[{{Salvaterra} {et~al}\mbox{.}(2011){Salvaterra}, {Ferrara}, \&
  {Dayal}}]{2011MNRAS.414..847S}
{Salvaterra} R., {Ferrara} A., {Dayal} P., 2011, \mnras, 414, 847

\bibitem[{{Salvaterra} {et~al}\mbox{.}(2006){Salvaterra}, {Magliocchetti},
  {Ferrara}, \& {Schneider}}]{2006MNRAS.368L...6S}
{Salvaterra} R., {Magliocchetti} M., {Ferrara} A., {Schneider} R., 2006,
  \mnras, 368, L6

\bibitem[{{Santos} {et~al}\mbox{.}(2002){Santos}, {Bromm}, \&
  {Kamionkowski}}]{2002MNRAS.336.1082S}
{Santos} M.~R., {Bromm} V., {Kamionkowski} M., 2002, \mnras, 336, 1082

\bibitem[{{Seo} {et~al}\mbox{.}(2015){Seo}, {Lee}, {Matsumoto}, {Jeong}, {Lee},
  \& {Pyo}}]{2015arXiv150405681S}
{Seo} H.~J., {Lee} H.~M., {Matsumoto} T., {Jeong} W.-S., {Lee} M.~G., {Pyo} J.,
  2015, ArXiv e-prints, 1504.05681

\bibitem[{{Thompson} {et~al}\mbox{.}(2007){Thompson}, {Eisenstein}, {Fan},
  {Rieke}, \& {Kennicutt}}]{2007ApJ...657..669T}
{Thompson} R.~I., {Eisenstein} D., {Fan} X., {Rieke} M., {Kennicutt} R.~C.,
  2007, \apj, 657, 669

\bibitem[{{V{\'a}zquez} \& {Leitherer}(2005)}]{2005ApJ...621..695V}
{V{\'a}zquez} G.~A., {Leitherer} C., 2005, \apj, 621, 695

\bibitem[{{Yue} {et~al}\mbox{.}(2015){Yue}, {Ferrara}, {Pallottini},
  {Gallerani}, \& {Vallini}}]{2015MNRAS.450.3829Y}
{Yue} B., {Ferrara} A., {Pallottini} A., {Gallerani} S., {Vallini} L., 2015,
  \mnras, 450, 3829

\bibitem[{{Yue} {et~al}\mbox{.}(2013{\natexlab{a}}){Yue}, {Ferrara},
  {Salvaterra}, \& {Chen}}]{2013MNRAS.431..383Y}
{Yue} B., {Ferrara} A., {Salvaterra} R., {Chen} X., 2013{\natexlab{a}}, \mnras,
  431, 383

\bibitem[{{Yue} {et~al}\mbox{.}(2013{\natexlab{b}}){Yue}, {Ferrara},
  {Salvaterra}, {Xu}, \& {Chen}}]{2013MNRAS.433.1556Y}
{Yue} B., {Ferrara} A., {Salvaterra} R., {Xu} Y., {Chen} X.,
  2013{\natexlab{b}}, \mnras, 433, 1556

\bibitem[{{Yue} {et~al}\mbox{.}(2014){Yue}, {Ferrara}, {Salvaterra}, {Xu}, \&
  {Chen}}]{2014MNRAS.440.1263Y}
{Yue} B., {Ferrara} A., {Salvaterra} R., {Xu} Y., {Chen} X., 2014, \mnras, 440,
  1263

\bibitem[{{Zemcov} {et~al}\mbox{.}(2014){Zemcov}, {Smidt}, {Arai}, {Bock},
  {Cooray}, {Gong}, {Kim}, {Korngut}, {Lam}, {Lee}, {Matsumoto}, {Matsuura},
  {Nam}, {Roudier}, {Tsumura}, \& {Wada}}]{2014Sci...346..732Z}
{Zemcov} M. {et~al.}, 2014, Science, 346, 732

\end{thebibliography}
 \end{document}